# Enhancement of a Novel Method for Mutational Disease Prediction using Bioinformatics Techniques and Backpropagation Algorithm

Ayad Ghany Ismaeel, Anar Auda Ablahad

**Abstract**— The noval method for mutational disease prediction using bioinformatics tools and datasets for diagnosis the malignant mutations with powerful Artificial Neural Network (Backpropagation Network) for classifying these malignant mutations are related to gene(s) (e.g. BRCA1 and BRCA2) cause a disease (breast cancer). This noval method didn't take in consideration just like adopted for dealing, analyzing and treat the gene sequences for extracting useful information from the sequence, also exceeded the environment factors which play important roles in deciding and calculating some of genes features in order to view its functional parts and relations to diseases. This paper is proposed an enhancement of a novel method as a first way for diagnosis and prediction the disease by mutations considering and introducing multi other features show the alternations, changes in the environment as well as genes, comparing sequences to gain information about the structure/function of a query sequence, also proposing optimal and more accurate system for classification and dealing with specific disorder using backpropagation with mean square rate 0.000000001.

———————— ◆ ————————

## 1 INTRODUCTION

CANCER is a dynamic disease, it has been estimated that a new genetic alteration occurs in one out of 10000 tumor cells at each cell division. This leads to an increasing number of genetic aberrations in cancer cells, some of which result in altered gene expression. These changes in gene expression are at least partially responsible for the characteristic of a tumor or tumor-type. Genome-wide gene expression profiles provide detailed "tumor signatures" and can potentially be used for molecular diagnosis and classification of tumors [1]. Cancer is a complex family of diseases, from the view of molecular biology; cancer is a genetic disease resulting from abnormal gene expression. This alternation of gene expression could be resulting from DNA instability, such as translocation, amplification, deletion or point mutations [2]. Breast cancer remains the most frequently diagnosed and the second most lethal cancer for women in the world[3]

The diagnosis and prediction of the cancer are important measures to avoid this disease specially for those which they have in their family [4], and that represented in novel method for mutational disease prediction as shown in Fig. 1, this noval method based on two approaches the first for diagnosis nucleotide and its protein mutations through bioinformatics tools.

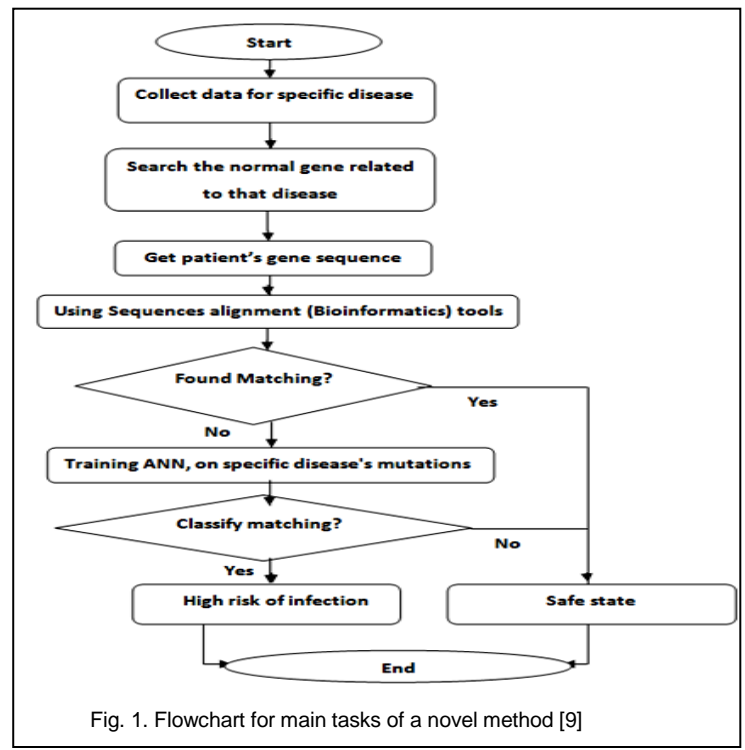

Fig. 1. Flowchart for main tasks of a novel method [9]

- *Ayad Ghany Ismaeel is currently pursuing Professor Assistant, PhD computer science in department of Information System Engineering, Erbil Technical College- Hawler Polytechnic University (previous FTE- Erbil), Iraq. PH-0096477035802999. E-mail: dr_a_gh_i@yahoo.com Alternative dr.ayad.ghany.ismaeel@gmail.com*
- *Anar Auda Ablahad is currently MSc student at research stage in computer science at Zakho University, its member in electrical and computer eng. Duhok-IRAQ. E-mail: anar_alkasyonan@yahoo.com*

And the second classify gene mutations (that have been diagnosed) are related to breast cancer or other cancer tissues using backpropagation neural network [9].

Important of Genome/Gene sequence in obtaining a blueprint–DNA directs all the instructions needed for cell development and function, DNA underlies almost every aspect of human health, in function and disfunction, to study gene expression in a specific tissue, organ or tumor, to study human variation, studying how humans relate to other organisms, finding correlations how genome information relates to development of cancer, susceptibility to certain diseases and drug metabolism is great challenge with providing tools for viewing, analyzing, discovering features and genes properties [5].

GC content is different in genes compared to the whole genome. In genes, the GC content is centered a round to 45-50%, and it is more uniformly distributed than in the genome. The genome is mostly 38% GC with its distribution skewed to the left. As a consequence, regions of high GC content (62-68%) have higher relative gene density than regions of lower GC content [6]. And also Mutator genes, especially those with dominant effects on the mutation spectra, are biased towards either GC or AT richness [7].

Homology search through The BLAST program works on the principle that regions of similarity are likely to contain strongly conserved, indel-resistant segments. These areas of strong conservation show up as non-gapped blocks in an alignment. Recent versions are also capable of producing gapped alignments, but they still build these gapped alignments up from non-gapped segments. This means that BLAST is most likely to fail to detect a homologous sequence where indels are scattered liberally and regularly throughout: in other words, highly divergent sequences may be missed. In a pairwise search, a query sequence is compared to a database sequence, yielding a score that indicates the likelihood of homology. This comparison is repeated for every sequence in the database and high-scoring hits are reported [8].

So there is multi features the noval method for mutational disease prediction must be taken in consideration like adopted for dealing, analyzing and treat the gene sequences for extracting useful information from the sequence, also the environment factor play important roles in deciding and calculating some of genes features in order to view its functional part and relations to diseases. By using bioinformatics tools and datasets with powerful Artificial Neural Network ANN (Back-Porpagation Network BPN) structure can make improvement/enhancement of a novel method.

## 2 RELATED WORK

Ayad Ghany Ismaeel, Anar A. Ablahad [9] developed a novel method for disease mutation prediction and classification based on two approaches which are first: diagnosis nucleotide and its protein mutations through bioinformatics tools (fasta, BioEdit and ClustalW). And second: classifying (BRCA1 and BRCA2) gene mutations whether these mutations (malignant) related to breast cancer or other cancer tissues using backpropagation neural network. The result of the paper showed reliable and accurate results with actual results of mutations diagnosis and classification.

Novel method reached to predict completely and obtained best results but there are some points not taken into consideration although are important for this noval method. Like the dataset adopted in the method and the normal gene used for comparison, also gene sequence analysis and gene features which play important roles in dealing and dis-covering gene actions and abilities.

The motivation overcome the drawbacks to reach an enhancement of a novel method, considering and introducing multi other features show the alternations, changes in the environment as well as genes, comparing sequences to gain information about the structure/function of a query sequence, also proposing optimal and more accurate system for classification and dealing with specific disorder.

## 3 PROPOSED AN ENHANCEMENT OF A NOVEL METHOD

The collection datasets relate to gene caused a disease is very important stage in the proposed of enhancement a noval method, because the noval method didn't care about it, however as show it can play an important role for mutational disease predication, as referring to section 1.

This proposal of enhancement a noval method focus on the important feature in Bioinformatics tools which is BLAST to reach to best Homology sequence related to the environment which is different from region to another.

The algorithm of main tasks for enhancement a novel method of maturational disease prediction, which based on overcome the drawbacks at novel method [9], this enhancement will refer to it in red as follow:

---

**Algorithm of main tasks for Enhancement a Novel method**

**Input:** DNA sequence (normal and person's gene sequence)

**Output:** Classification of tumor gene mutations (for the patient)

---

**BEGIN**

   **Step 1:** collect datasets for the project

   **Step 2:** search normal Homology Sequences at BLAST

   **Step 3:** calculate GC% and AT%

   **Step 4:** Find the percentage of normal gene homology e.g. at NCBI

   **Step 5:** if GC% satisfies 38% then

            Depend it

         Else find another normal Homology seguence form another

            Database like EBI, Ensemble, etc and return to step 2

  **Step 6:** Get normal gene sequences

   **Step 7:** CREATE fasta file;

         SET normal and Person's gene sequence to fasta file

   **Step8:** SET fasta file to clustalW for alignment; Display alignment result

   **Step9: Simulate ANN (BPN) of optimal (new) structure**.

   **Step10: CREATE training dataset file of Breast cancer mutations**

         Input DNA gene sequence (of the patient);

         Display BPN results;

         If matching found then

            Display"highly risk of breast cancer"

         Else Disply no risk "Normal"

**END**

## 4 EXPERIMENTAL RESULTS

### 4.1 Bioinformatics Techniques

Implement the proposed enhancement of noval method will apply to diagnosis the mutations at patient gene and its protein is malignant or not.

In order to know how to access and get helpful bioinformatics tools this modified approach to reach an enhancement of a novel method consider the collection dataset related to gene caused the disease is very important stage in the proposed of enhancement a noval method, because the noval method which not care about it for mutational disease predication, so in this stage will consider additional parts as follow:

1. The data set adapted to depend on the environment, as it plays roles in alternations and differences in genes shapes also genes mutations.
2. The researchers or biologists adopt a normal gene for analysis and comparison should be compatible with the environment of the study. The improvement of this point can be shown in Fig. 2. The normal gene can be obtained from NCBI or EBI, Ensemble, COSMIC. and each of these databases provide different form of normal gene adopted in the local database but when search for homology sequences at NCBI, the 100% match found between the two normal gene sequences. When normal gene of COSMIC is entered to NCBI at blast it shows the RefSeq (normal) as the first one in homology list with max identities=100.

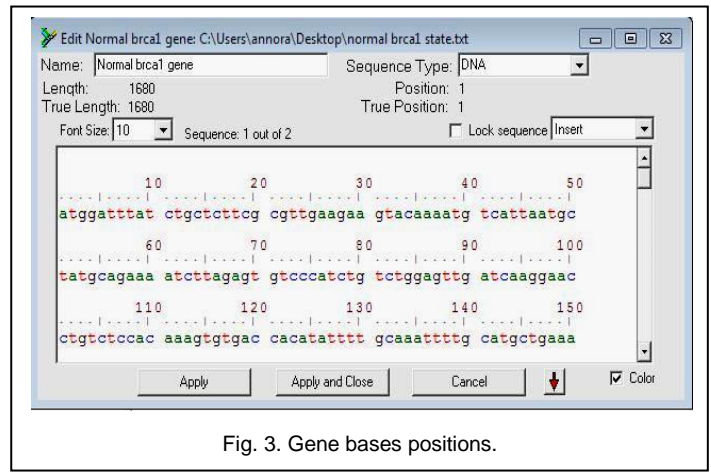

Fig. 3. Gene bases positions.

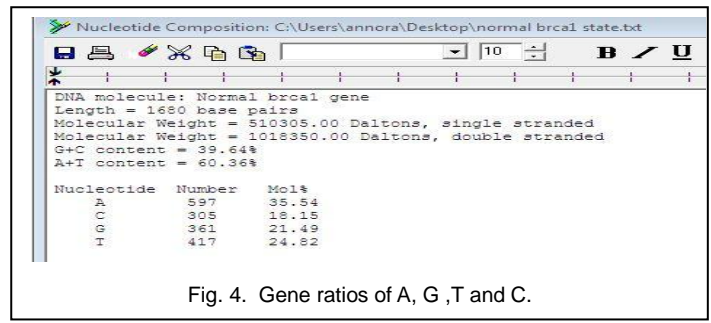

Fig. 4. Gene ratios of A, G ,T and C.

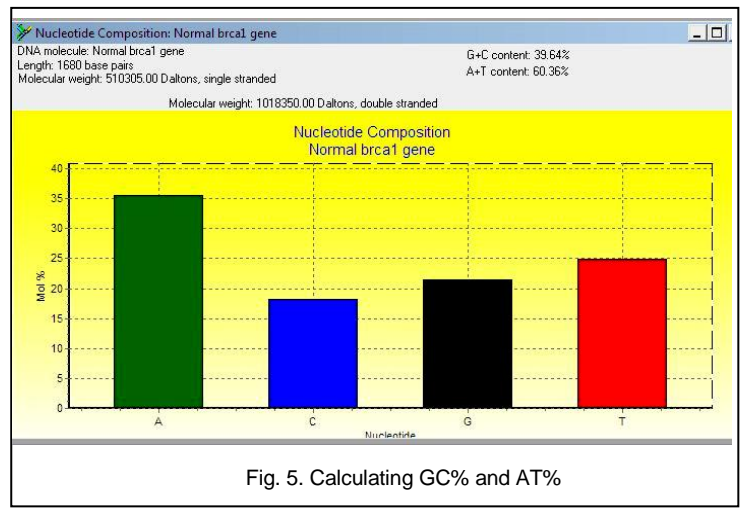

Fig. 5. Calculating GC% and AT%

3. Calculating GC%content and AT% content for gene stability and ability of doing amplification can be done by using BioEdit package to determine ratios and some amino acid and protein features. Needed to calculate GC% and AT% of normal BRCA1 gene to determine is the genome around 38% GC to depend it otherwise search about another normal BRCA1 gene form another database, shown in Fig. 3, Fig. 4 and Fig. 5 respectively.

Fig. 2. Adopt a normal gene for analysis and comparison.

When reaching to this stage will need to complete another important bioinformatics tool which is ClustalW, that tool can work with fasta file for diagnosis there are malignant mutations or not, as referring to it in noval method [9].

### 4.2 Backpropagation Algorithm

To predict these malignant mutations which obtained in subsection 4.1 (using bioinformatics tools) are related to BRCA1 and BRCA2 (which caused breast cancer) based on noval method need training the BPN by all malignant mutations (which are 5 related to BRCA1 and 4 related to BRCA2) then used to classify these malignant mutations which caused breast cancer [9].

The new modification in the BPN, i.e. get a new structure which can be summarized as the optimum and accurate result with best performance comparing to the previous novel method, the BPN is most commonly used in medical research. BPNs where the signals travel in one direction from input neuron to an output neuron without returning to its source, neural network is a sorted topology which include node and weighted connections. Each layer inside neural network will be connected by their weight connector; the new BPN consists of (input layer, at least one hidden layer and output layer) as shown in Fig. 6 [6].

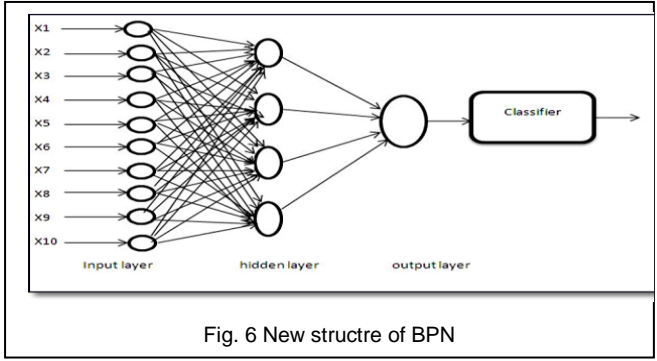

Fig. 6 New structre of BPN

That means the enhancement of a novel method will be used a feed forward back-propagation network as popular model in neural network, the new structure of BPN consist of:
1. One input layer with 10 nodes at the input layer
2. One hidden layer with 4 nodes at hidden layer
3. One output layer with one node at the output layer.

Using MATLAB R2010a on PC type Core i3 for neural network toolbox because it contains various functions the classification of malignant mutations for breast cancer was developed successfully using new structure of feed-forward BP neural network to obtain an optimal classifier for prediction a breast cancer or any other disease with mean square error 0.000000001, as shown in Fig. 7.

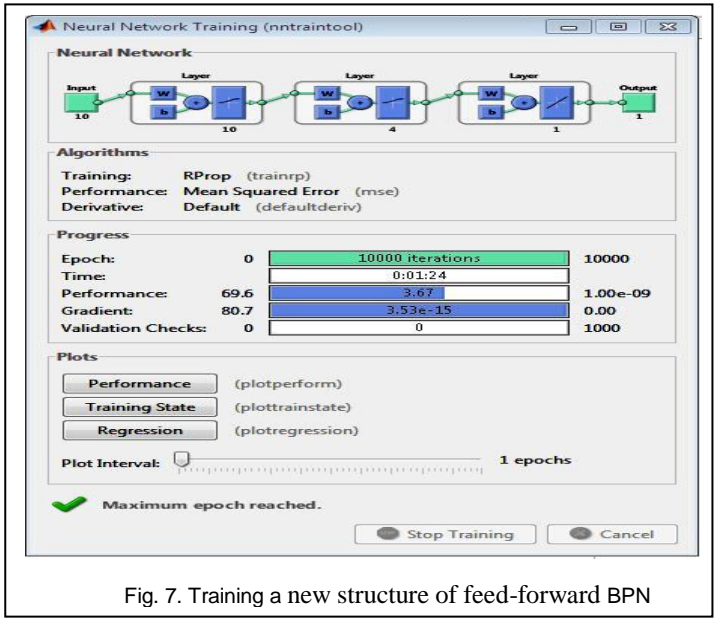

Fig. 7. Training a new structure of feed-forward BPN

Fig. 8 reveals plot of the important element of training new structure of feed-forward BPN, which is the plot of performance.

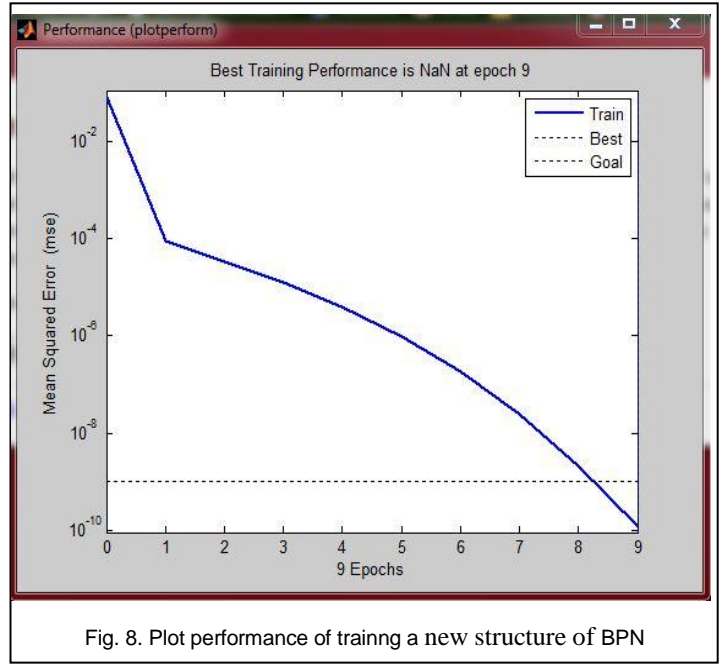

Fig. 8. Plot performance of trainng a new structure of BPN

### 4.3 Discussion the Results

Table 1 reveals comparing the results of implementing an enhancement of a novel method with the noval method (the origin) for mutational disease prediction [9].

TABLE 1
REVEALS COMPARISON AN ENHANCEMENT OF A NOVEL METHOD WITH NOVEL METHOD FOR MUTATIONAL DISEASE PREDICTION

| The Feature | Enancement of a Novel Method | Novel Method (the origin )[9] |
|---|---|---|
| Datasets adopted is depended of the environment, as it play roles in alternations and differences in genes shapes and genes mutations. | Yes | No |
| Normal gene can be obtained different form adopted to the local database (based on their environment) | Yes | No |
| Considered the GC% content and AT% content for gene | Yes | No |
| Mean square rate. | Mean square rate 0.000000001 | Mean square rate 0.0000001 |
| Optimize the classifying BPN. | Yes | No |

## 5 CONCLUSIONS

The important conclusions which obtained from implementing the proposed enhancement of novel method for mutational disease prediction are:

1. The enhancement of a novel method will base on collection datasets which related to environment, so that means the results of diagnosis using bioinformatics tools and then the classifying of BPN, i.e. the prediction is more expressive for the region and will be better as this as have seen from the above impact on the results and how it could affect the environment.

2. This enhancement of a novel method based on calculating the GC% content and AT% content for selecting the a normal gene caused the disease and make GC%= 38% as measure to depend the normal Homology sequence which is playing important role to select the appropriate normal Homology sequence related to environment.

3. Offers an automatic and friendly diagnosis system for detecting malignant mutation and pre-predict of breast cancer as shown in Table 1, i.e. can use by any researcher or patient who needed to test malignant mutations at gene(s) which caused any cancer known its malignant mutations.

4. This Enhancement model of classification malignant mutations for breast cancer was developed successfully using new structure of feed-forward back-propagation neural network, i.e. obtained a best classifier for prediction a breast cancer or any other disease disease known its malignant mutations with mean square error 0.000000001.